# An Enhanced Leakage-Based Precoding Scheme for Multi-User Multi-Layer MIMO Systems

C. Yang

*Abstract*—In this letter, we propose an enhanced leakage-based precoding scheme, i.e., layer signal to leakage plus noise ratio (layer SLNR) scheme, for multi-user multi-layer MIMO systems. Specifically, the layer SLNR scheme incorporates the MIMO receiver structure into the precoder design procedure, which makes the formulation of signal power and interference / leakage power more accurate. Besides, the layer SLNR scheme not only takes into account the inter-layer interference from different users, but also takes care of the inter-layer interference from the same user which is usually assumed to be zero in previous studies. As a result, the proposed layer SLNR scheme produces a good balance between the layer signal power and layer interference power in a multi-user multi-layer MIMO system, therefore achieves better system performance. The effectiveness and superiority of the proposed layer SLNR scheme are validated via simulations.

*Index Terms*—Multi-user multi-layer MIMO, layer SLNR.

## I. INTRODUCTION

WITH the fast development of mobile internet, the demand for high data rate is growing explosively. Multi-antenna precoding techniques are effective ways of increasing data rate. Thus, they have been widely studied and applied in current wireless communication systems. Among those, the leakage-based (or signal to leakage plus noise ratio, SLNR) precoding scheme firstly proposed in [1] is especially effective in balancing the desired signal power and the leakage / interference power in multi-user MIMO systems. Besides, a simple closed-form solution can be obtained by using this leakage-based precoding scheme. Therefore, it has attracted lots of interest both from the academia and the industry.

The SLNR-based schemes have been extensively studied in multi-user MIMO systems [2]-[6]. An active antenna selection scheme is proposed based on the SLNR criterion in which high geometry (high signal to interference ratio, high SIR) users reduce the number of active receive antennas to "help" the low geometry (low SIR) users to achieve their target SIRs [2]. A generation of SLNR precoding to MIMO orthogonal frequency division multiplexing (OFDM) multi-user systems is studied in [3]. In [4], a power allocation scheme combining the interference and noise whitening, the singular value decomposition and the water-filling algorithms is investigated for a max-SLNR precoder per user. A new linear precoding scheme by slightly relaxing the SLNR maximization for multi-user MIMO systems with multiple data streams per user is proposed in [5]. Their scheme reduces the gap between the per-stream effective channel gains which is inherent limitations in the original SLNR precoding scheme. In [6], the authors derive equivalent expressions of SLNR-based precoding solutions and establish the equivalence between the SLNR, the regularized block diagonalization, and the generalized minimum mean square error (MMSE) channel inversion method 2 precoding schemes. The performance of the SLNR scheme is also analyzed basing on this equivalent form.

As we can see from the above, although promising results have been shown and many useful theoretical results have been derived in previous studies, the original SLNR scheme and its variants are not yet perfect, and therefore have room for performance improvement. Specifically, just as having been pointed out in [7], those previous leakage-based schemes do not take into account the various MIMO receiver structure in designing the precoders. Thus, the modeling of signal power and interference / leakage power is not accurate, or, the degree of freedom offered by the receiver is not fully exploited in the precoder design. Besides, the previous schemes do not consider the inter-layer interference from the same user which may have significant impact on the performance in a multi-layer scenario. They roughly assume no inter-layer interference from the same user and define the SLNR on a per-user basis, which leads to the lack of capability of balancing layer signal power and layer interference power in a multi-user multi-layer scenario.

In this letter, we analyze the limitations of the original SLNR-based schemes in multi-layer MIMO systems and propose an enhanced SLNR scheme, termed layer SLNR, which not only takes into account the MIMO receiver structure, but also defines the SLNR on a per-layer basis. Therefore, the layer SLNR scheme formulizes the inter-layer interference from different users as well as from the same user. We found that the proposed layer SLNR scheme in this letter can be seen as a generation of the scheme in [7] which only considers a single-layer per user scenario to a multi-layer per user scenario.

The remainder of this letter is as follows. In Section II, we analyze the original SLNR-based precoding scheme for a multi-user multi-layer scenario, and address their limitations. Then an enhanced scheme (i.e., layer SLNR) is proposed in Section III. And the simulation results are provided to validate the effectiveness and superiority of the layer SLNR scheme in Section IV. Conclusions are given in Section V.

## II. THE ANALYSIS OF THE ORIGINAL SLNR SCHEME FOR MULTI-USER MULTI-LAYER MIMO SYSTEMS

In this section, we review the original SLNR scheme in a multi-user multi-layer MIMO system, where a downlink single base station and multi-user ($K$ users) scenario with $N$ transmit antennas at the base station and $M_k$ receive antennas at the $k$th user is depicted in Fig. 1. Let $\mathbf{s}_k(t) \in \mathbb{C}^{L_k \times 1}$ denote the data symbol vector to be transmitted to the $k$th user at time $t$, where $L_k$ is the number of data layers of user $k$ ($1 \leq k \leq K$), and $\mathbb{E}\left[\mathbf{s}_k(t)\mathbf{s}_k^H(t)\right] = (1/L_k)\mathbf{I}$. The data symbol vector $\mathbf{s}_k(t)$ is then multiplied by an $N \times L_k$ precoding matrix $\mathbf{V}_k(t)$ before being transmitted in the wireless channel, where $\text{Tr}\left(\mathbf{V}_k^H(t)\mathbf{V}_k(t)\right) = L_k$, and $\text{Tr}(\mathbf{X})$ is the trace of matrix $\mathbf{X}$.

Assuming a narrowband flat MIMO channel (e.g., a subcarrier of an OFDM system) with perfect synchronization, the baseband, equivalent received signal of the $k$th user at time $t$ can be written as follows,

$$\mathbf{y}_k(t) = \mathbf{H}_k(t)\sum\nolimits_{i=1}^{K}\mathbf{V}_i(t)\mathbf{s}_i(t) + \mathbf{w}_k(t), \quad (1)$$

where $\mathbf{H}_k(t)$ is an $M_k \times N$ MIMO channel matrix from the base station transmit antennas to the receive antennas of user $k$ at time $t$. Each entry in $\mathbf{H}_k(t)$ is modeled by an independent and identical complex Gaussian variable with zero-mean and unit-variance. $\mathbf{w}_k(t)$ is a length $M_k$ noise vector at user $k$ at time $t$. We assume that the entries of $\mathbf{w}_k(t)$ are independent and identically distributed according to $\mathcal{CN}(0, \sigma_k^2(t))$ and are spatially white, i.e., $\mathbb{E}\left[\mathbf{w}_k(t)\mathbf{w}_k^*(t)\right] = \sigma_k^2(t)\mathbf{I}$. For notation simplicity, we omit the time index $t$ in the following discussions when there is no confusion.

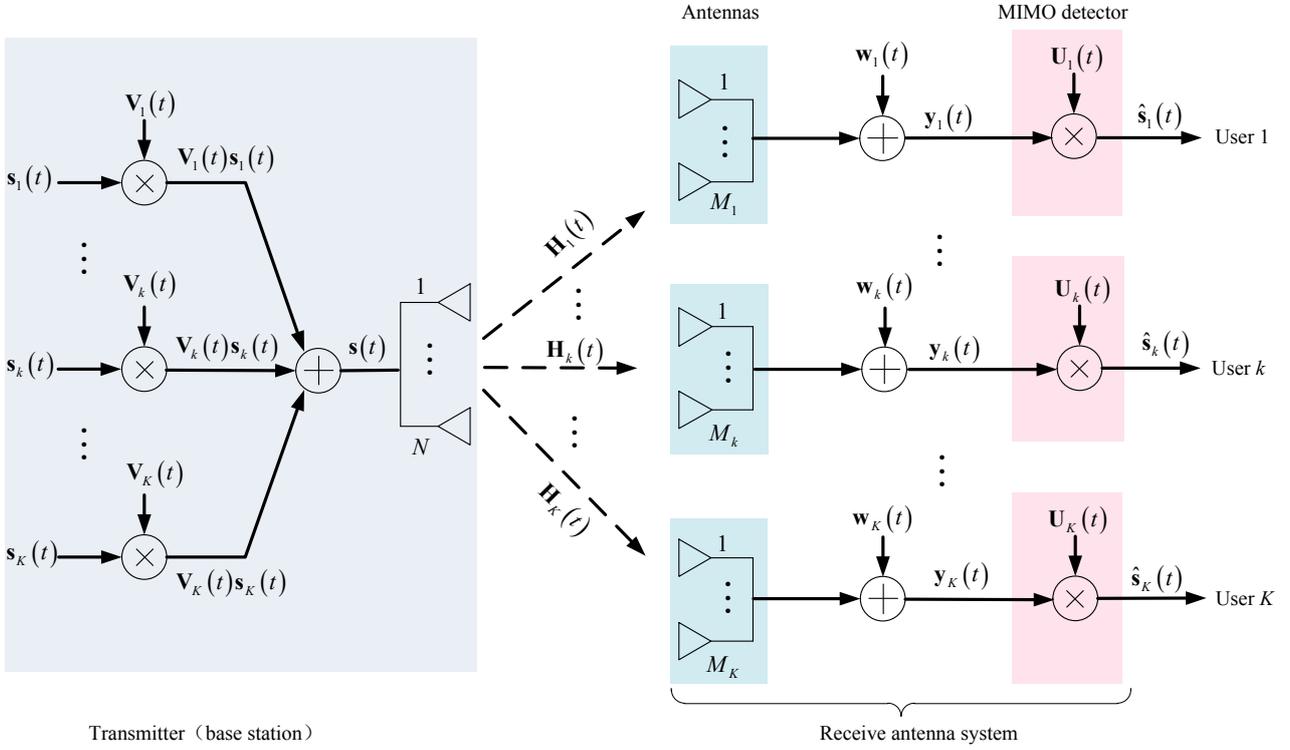

Fig. 1 The multi-user multi-layer MIMO system block diagram.

According to [1], the SLNR for user $k$ is defined as follows,

$$\mathrm{SLNR}_k = \frac{(1/L_k)\|\mathbf{H}_k \mathbf{V}_k\|_{\mathrm{F}}^2}{M_k \sigma_k^2 + \sum_{i=1, i\neq k}^{K}(1/L_k)\|\mathbf{H}_i \mathbf{V}_k\|_{\mathrm{F}}^2} \\ = \frac{\mathrm{Tr}\left(\mathbf{V}_k^H \mathbf{H}_k^H \mathbf{H}_k \mathbf{V}_k\right)}{\mathrm{Tr}\left(\mathbf{V}_k^H \left(M_k \sigma_k^2 \mathbf{I} + \tilde{\mathbf{H}}_k^H \tilde{\mathbf{H}}_k\right)\mathbf{V}_k\right)}, \quad (2)$$

where $\tilde{\mathbf{H}}_k \triangleq [\mathbf{H}_1 \cdots \mathbf{H}_{k-1}\ \mathbf{H}_{k+1} \cdots \mathbf{H}_K]^T$ is the leakage / interference channel corresponding to user $k$, and $\|\mathbf{X}\|_{\mathrm{F}}$ is the Frobenius norm of matrix $\mathbf{X}$.

The transmit precoder targeted for user $k$ that maximizes its SLNR is given by

$$\hat{\mathbf{V}}_k = \arg\max_{\mathbf{V}_k \in \mathbb{C}^{N \times L_k}} \mathrm{SLNR}_k \\ = \arg\max_{\mathbf{V}_k \in \mathbb{C}^{N \times L_k}} \frac{\mathrm{Tr}\left(\mathbf{V}_k^H \mathbf{H}_k^H \mathbf{H}_k \mathbf{V}_k\right)}{\mathrm{Tr}\left(\mathbf{V}_k^H \left(M_k \sigma_k^2 \mathbf{I} + \tilde{\mathbf{H}}_k^H \tilde{\mathbf{H}}_k\right)\mathbf{V}_k\right)}. \quad (3)$$

Thus, the optimal precoder for user $k$ is shown to be the generalized eigenvectors of matrix pair $\mathbf{H}_k^H \mathbf{H}_k$ and $M_k \sigma_k^2 \mathbf{I} + \tilde{\mathbf{H}}_k^H \tilde{\mathbf{H}}_k$ corresponding to the leading $L_k$ eigenvalues.

From the above definition of the original SLNR and the derivation in Section V in [1], we can observe that: 1) for the desired signal power, i.e., $(1/L_k)\|\mathbf{H}_k \mathbf{V}_k\|_{\mathrm{F}}^2$ in (2), the precoder design in the original SLNR scheme is based on the assumption that the structure of the MIMO detector is a matched filter type. Thus, there is no inter-layer interference within a user when a matched filter MIMO detector is employed by the user. 2) For the interference / leakage power, i.e., $\sum_{i=1, i\neq k}^{K}(1/L_k)\|\mathbf{H}_i \mathbf{V}_k\|_{\mathrm{F}}^2$ in (2), just as being pointed out in [7], the receiver structures of the interfered users are not considered in the original SLNR scheme; therefore, the modeling of interference caused by user $k$ to other co-channel users is not accurate, which leads to less effective of interference reduction by the designed precoder.

Therefore, similar to the single-layer scenario in [7], we incorporate the MIMO receiver structure into the original SLNR scheme. That is,

$$\mathrm{SLNR}_k = \frac{(1/L_k)\|\mathbf{U}_k \mathbf{H}_k \mathbf{V}_k\|_{\mathrm{F}}^2}{M_k \sigma_k^2 + \sum_{i=1, i\neq k}^{K}(1/L_k)\|\mathbf{U}_i \mathbf{H}_i \mathbf{V}_k\|_{\mathrm{F}}^2}. \quad (4)$$

Let's partition the precoding matrix $\mathbf{V}_k$ by column, i.e., $\mathbf{V}_k = [\mathbf{v}_{k,1}\ \mathbf{v}_{k,2}\ \cdots\ \mathbf{v}_{k,L_k}]$, and partition the MIMO receiver structure $\mathbf{U}_k$ by row, i.e., $\mathbf{U}_k = [\mathbf{u}_{k,1}\ \mathbf{u}_{k,2}\ \cdots\ \mathbf{u}_{k,L_k}]^T$, where $\mathbf{v}_{k,l}$ is the $l$th column of $\mathbf{V}_k$ and $\mathbf{u}_{k,l}$ is $l$th row of $\mathbf{U}_k$.

Therefore, the expression $\mathbf{U}_k \mathbf{H}_k \mathbf{V}_k$ in the numerator of (4) can be rewritten as follows,

$$\begin{bmatrix} \mathbf{u}_{k,1}\mathbf{H}_k\mathbf{v}_{k,1} & \cdots & \mathbf{u}_{k,1}\mathbf{H}_k\mathbf{v}_{k,l} & \cdots & \mathbf{u}_{k,1}\mathbf{H}_k\mathbf{v}_{k,L_k} \\ \vdots & \ddots & \vdots & \ddots & \vdots \\ \mathbf{u}_{k,l}\mathbf{H}_k\mathbf{v}_{k,1} & \cdots & \mathbf{u}_{k,l}\mathbf{H}_k\mathbf{v}_{k,l} & \cdots & \mathbf{u}_{k,l}\mathbf{H}_k\mathbf{v}_{k,L_k} \\ \vdots & \ddots & \vdots & \ddots & \vdots \\ \mathbf{u}_{k,L_k}\mathbf{H}_k\mathbf{v}_{k,1} & \cdots & \mathbf{u}_{k,L_k}\mathbf{H}_k\mathbf{v}_{k,l} & \cdots & \mathbf{u}_{k,L_k}\mathbf{H}_k\mathbf{v}_{k,L_k} \end{bmatrix} \in \mathbb{C}^{L_k \times L_k}. \quad (5)$$

Given the precoding matrix $\mathbf{V}_k$, matrix (5) exhibits different forms when different receiver structures are employed. For example, when the receiver structure is a matched filter type, i.e., $\mathbf{U}_k = (\mathbf{H}_k \mathbf{V}_k)^H / \|\mathbf{H}_k \mathbf{V}_k\|_{\mathrm{F}}$, matrix (5) is a diagonal matrix, i.e., all the off-diagonal elements in matrix (5) are zero. That is, in this case different data layers from the same user $k$ will not interfere with each other. Thus, the numerator $(1/L_k)\|\mathbf{U}_k\mathbf{H}_k\mathbf{V}_k\|_F^2$ in (4) reflects the power of the desired signal including multiple data layers (or sum of signal power of different data layers).

However, when the receiver structure is not a matched filter type, matrix (5) will not be a diagonal matrix in general. Therefore, the value of $(1/L_k)\|\mathbf{U}_k\mathbf{H}_k\mathbf{V}_k\|_F^2$ not only includes the power of desired signal (the sum of signal power of multiple data layers) of user $k$, but also includes the inter-layer interference power of user $k$ (the off-diagonal elements in matrix (5) are the inter-layer interference of user $k$). For example, $\mathbf{u}_{k,d}\mathbf{H}_k\mathbf{v}_{k,l}$ is non-zero interference from the $l$th data layer to the $d$th data layer of user $k$ for $l \neq d$.

Therefore, the definition in (4) is not general. This means that, by simply incorporating the receiver structure into the original SLNR scheme as having done in a single-layer scenario in [7] is now not general and accurate in a multi-layer scenario.

Actually, the definitions in (2) and in (4) are both on a per-user basis instead of on a per-layer basis. In a single-layer scenario, the definition on a per-user basis and that on a per-layer basis are the same. This is because each user has only one data layer. However, in a multi-layer scenario (each user has multiple data layers), the rationality and effectiveness of the definition of SLNR on a per-user basis needs further investigations. As a matter of fact, the rationality of the definition of SLNR on a per-user basis lies in the assumption that the MIMO detector of each user is a matched filter type as in [1], [5]. In this case, there is no inter-layer interference among the data layers from the same user. Therefore, it is reasonable to define the SLNR on a per-user basis from the viewpoint of desirable signal power.

However, in general, taking the cost, complexity, performance and other factors into consideration, users may employ different types of MIMO detectors like matched filters, MMSE filters, etc. Thus, the interference among different data layers from the same user is in general not zero. Therefore, the definition of SLNR on a per-user basis which assumes no inter-layer interference from the same user is not general. In the next section, we will propose an enhanced SLNR scheme that is general for any types of receiver structure and formulizes the inter-layer interference both from the same user and from different users.

### III. THE LAYER SLNR SCHEME

In this section, we propose an enhanced SLNR scheme, i.e., layer SLNR, which is general for any types of receiver structure. That is, define the SLNR for each data layer considering the inter-layer interference both from the same user and from different users, and design the precoder for each data layer no matter whether the data layers are from the same user or from different users.

More specifically, the definition of layer SLNR, i.e., the SLNR of the $l$th data layer from user $k$, is as follows,

$$\text{SLNR}_k^l = \frac{\left|[\mathbf{U}_k\mathbf{H}_k\mathbf{V}_k]_{l,l}\right|^2}{M_k\sigma_k^2 + \sum_{\substack{d=1\\d\neq l}}^{L_k}\left|[\mathbf{U}_k\mathbf{H}_k\mathbf{V}_k]_{d,l}\right|^2 + \sum_{\substack{i=1\\i\neq k}}^{K}\sum_{d=1}^{L_i}\left|[\mathbf{U}_i\mathbf{H}_i\mathbf{V}_k]_{d,l}\right|^2}, \quad (6)$$

where $[\mathbf{X}]_{d,l}$ is the element of matrix $\mathbf{X}$ located at the $d$th row and the $l$th column.

The numerator $\left|[\mathbf{U}_k\mathbf{H}_k\mathbf{V}_k]_{l,l}\right|^2$ in (6) denotes the received signal power of the $l$th data layer of user $k$. The expression $\sum_{\substack{d=1\\d\neq l}}^{L_k}\left|[\mathbf{U}_k\mathbf{H}_k\mathbf{V}_k]_{d,l}\right|^2$ in the denominator of (6) denotes the interference power caused by the $l$th data layer to other $L_k - 1$ data layers of user $k$. The expression $\sum_{\substack{i=1\\i\neq k}}^{K}\sum_{d=1}^{L_i}\left|[\mathbf{U}_i\mathbf{H}_i\mathbf{V}_k]_{d,l}\right|^2$ in the denominator of (6) denotes the interference power caused by the $l$th data layer of user $k$ to the data layers of other $K-1$ co-channel users.

According to the partitions of $\mathbf{V}_k$ and $\mathbf{U}_k$, (6) can be rewritten as follows,

$$\text{SLNR}_k^l = \frac{\left|\mathbf{u}_{k,l}\mathbf{H}_k\mathbf{v}_{k,l}\right|^2}{M_k\sigma_k^2 + \sum_{\substack{d=1\\d\neq l}}^{L_k}\left|\mathbf{u}_{k,d}\mathbf{H}_k\mathbf{v}_{k,l}\right|^2 + \sum_{\substack{i=1\\i\neq k}}^{K}\left\|\mathbf{U}_i\mathbf{H}_i\mathbf{v}_{k,l}\right\|_F^2}. \quad (7)$$

For the notation simplicity, we introduce the following notation definitions,

$$\mathbf{G}_{k,l} \triangleq \mathbf{u}_{k,l}\mathbf{H}_k, \quad \hat{\mathbf{U}}_{k,l} \triangleq \begin{bmatrix} \mathbf{u}_{k,1} & \cdots & \mathbf{u}_{k,(l-1)} & \mathbf{u}_{k,(l+1)} & \cdots & \mathbf{u}_{k,L_k} \end{bmatrix}^T,$$

$$\hat{\mathbf{G}}_{k,l} \triangleq \hat{\mathbf{U}}_{k,l}\mathbf{H}_k, \quad \breve{\mathbf{G}}_{k,l} \triangleq \begin{bmatrix} \mathbf{U}_1\mathbf{H}_1 & \cdots & \mathbf{U}_{k-1}\mathbf{H}_{k-1} & \mathbf{U}_{k+1}\mathbf{H}_{k+1} & \cdots & \mathbf{U}_K\mathbf{H}_K \end{bmatrix}^T,$$

where $\mathbf{G}_{k,l}$ is the effective desired channel of the $l$th data layer of user $k$, $\hat{\mathbf{G}}_{k,l}$ is effective interference / leakage channel of the $l$th data layer of user $k$ to other $L_k - 1$ data layers of user $k$, and $\breve{\mathbf{G}}_{k,l}$ is the effective interference / leakage channel of the $l$th data layer of user $k$ to the data layers of other $K - 1$ co-channel users.

Based on the above notation definitions, (7) can be rewritten as,

$$\text{SLNR}_k^l = \frac{\left|\mathbf{G}_{k,l}\mathbf{v}_{k,l}\right|^2}{M_k\sigma_k^2 + \left\|\hat{\mathbf{G}}_{k,l}\mathbf{v}_{k,l}\right\|_F^2 + \left\|\breve{\mathbf{G}}_{k,l}\mathbf{v}_{k,l}\right\|_F^2}.$$

Let $\bar{\mathbf{G}}_{k,l} \triangleq \begin{bmatrix} \hat{\mathbf{G}}_{k,l} & \breve{\mathbf{G}}_{k,l} \end{bmatrix}^T$, then

$$\text{SLNR}_k^l = \frac{\left|\mathbf{G}_{k,l}\mathbf{v}_{k,l}\right|^2}{M_k\sigma_k^2 + \left\|\bar{\mathbf{G}}_{k,l}\mathbf{v}_{k,l}\right\|_F^2}$$

$$= \frac{\text{Tr}\left(\mathbf{v}_{k,l}^H\mathbf{G}_{k,l}^H\mathbf{G}_{k,l}\mathbf{v}_{k,l}\right)}{\text{Tr}\left(\mathbf{v}_{k,l}^H\left(M_k\sigma_k^2\mathbf{I} + \bar{\mathbf{G}}_{k,l}^H\bar{\mathbf{G}}_{k,l}\right)\mathbf{v}_{k,l}\right)}.$$

Therefore, the precoder for the $l$th data layer of user $k$ can be obtained via the following formula,

$$\hat{\mathbf{v}}_{k,l} = \arg\max_{\mathbf{v}_{k,l}\in\mathbb{C}^{N\times 1}}\text{SLNR}_k^l$$

$$= \arg\max_{\mathbf{v}_{k,l}\in\mathbb{C}^{N\times 1}}\frac{\text{Tr}\left(\mathbf{v}_{k,l}^H\mathbf{G}_{k,l}^H\mathbf{G}_{k,l}\mathbf{v}_{k,l}\right)}{\text{Tr}\left(\mathbf{v}_{k,l}^H\left(M_k\sigma_k^2\mathbf{I} + \bar{\mathbf{G}}_{k,l}^H\bar{\mathbf{G}}_{k,l}\right)\mathbf{v}_{k,l}\right)}. \quad (8)$$

That is, the optimal precoder is the generalized eigenvector corresponding to the maximum eigenvalue of the matrix pair $\mathbf{G}_{k,l}^H\mathbf{G}_{k,l}$ and $M_k\sigma_k^2\mathbf{I} + \bar{\mathbf{G}}_{k,l}^H\bar{\mathbf{G}}_{k,l}$.

Note that the above proposed layer SLNR scheme can be seen as a generation of the scheme in [7] to a multi-layer scenario. That is, when each user has a single layer data, the proposed layer SLNR scheme is the same as the scheme in [7].

### IV. SIMULATION RESULTS

In this section, we validate the proposed layer SLNR precoding scheme via simulations. The performance of the original SLNR in [1] was also shown as a comparison. The simulations were performed on a MIMO-OFDM system simulator in which the number of antennas at the base station ( or transmitter) was eight ( $N = 8$ ) or ten ( $N = 10$ ), and the number of users was three ( $K = 3$ ), each of which was equipped with three receive antennas ( $M_k = 3$, $k = 1,2,3$ ), and the number of data layers for each user was two ( $L_k = 2$, $k = 1,2,3$ ). The SNR per receive antenna is $1/\sigma^2 = 0\,\text{dB}$. In the simulations, all users employ matched filter MIMO receivers as in [1], i.e., $\mathbf{U}_k = (\mathbf{H}_k \mathbf{V}_k)^H / \|\mathbf{H}_k \mathbf{V}_k\|$. Then the simulations with a different MIMO receiver structure, i.e., the multi-user MMSE MIMO receiver $\mathbf{U}_k = (\mathbf{H}_k \mathbf{V}_k)^H ((\mathbf{H}_k \mathbf{V}_k)(\mathbf{H}_k \mathbf{V}_k)^H + \mathbf{R}_k)^{-1}$, were also performed, where $\mathbf{R}_k = \sigma_k^2 \mathbf{I} + \sum_{i=1, i \neq k}^{K} (\mathbf{H}_i \mathbf{V}_i)(\mathbf{H}_i \mathbf{V}_i)^H$.

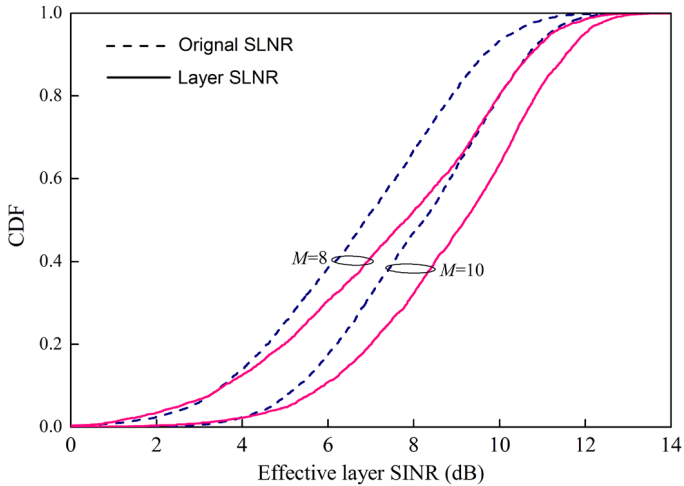 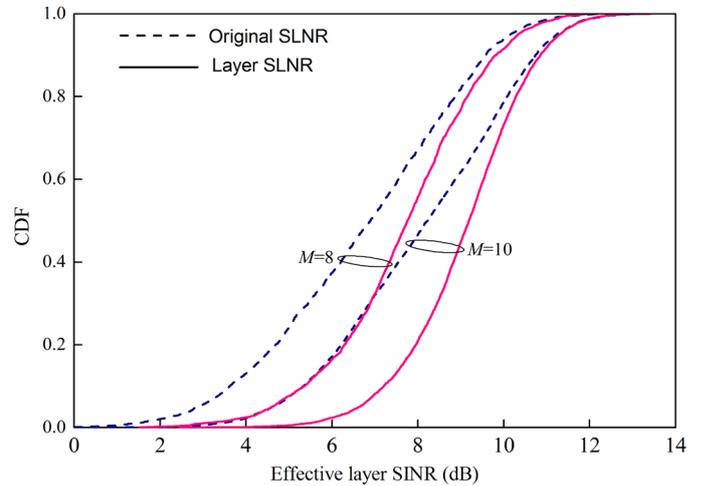

Fig. 2. CDFs of the effective layer SINR for different SLNR-based precoding schemes with a matched filter type MIMO detector.

Fig. 3. CDFs of the effective layer SINR for different SLNR-based precoding schemes with an MMSE type MIMO detector.

In the simulations, $\mathbf{U}_k(t-1)$, the latest MIMO receiver structure information available, was used as an approximation to $\mathbf{U}_k(t)$. As a result, user $k$ fed back $\mathbf{U}_k(t-1)\mathbf{H}_k(t)$ (i.e., the augmented $\mathbf{H}_k(t)$), as opposed to $\mathbf{H}_k(t)$ in the original SLNR precoding, to the transmitter. The feedback overhead is thus the same as in [1]. Initially, when $\mathbf{U}_k(t-1)$ was not available, e.g., at the very beginning of the transmission ( $t = 0$ ), the original SLNR solution (3) was used. For $t > 0$, (8) was then used.

Fig. 2 and Fig. 3 show the cumulative distribution function (CDF) curves of the effective layer signal to interference plus noise ratio (SINR) with a matched filter MIMO detector and an MMSE MIMO detector, respectively. Note that the metric "effective layer SINR" instead of the metric "effective user SINR" is used. This is because "effective user SINR" cannot reflect the performance accurately. For example, two users who have the same "effective user SINR" may have quite different "effective layer SINR"s in a multi-layer per user scenario. From Fig. 2 and Fig. 3, we can observe that: 1) the performance of the proposed layer SLNR scheme is about 1 dB better than that of the original SLNR scheme in median and high SINR regions when a matched filter type MIMO detector is used; 2) the performance of the proposed layer SLNR scheme is about 1.5 dB better than that of the original SLNR scheme in low and median SINR regions when an MMSE type MIMO detector is used. The reason is as follows. 1) In the layer SLNR scheme, the receiver structure is taking into account in designing precoders, which describes the signal and interference power more accurately; therefore, better performance is achieved. 2) The layer SLNR scheme is defined on a per-layer basis, which not only balances the inter-layer interference from different users but also balances the inter-layer interference from the same user; therefore, for each data layer, the designed precoder achieves a good balance between interference power and signal power, thus smaller layer performance difference within a user is achieved.

## V. CONCLUSION

Among a wide variety of MIMO precoding schemes, the leakage-based precoding scheme is especially popular / promising and attractive in reducing co-channel interference in multi-user MIMO systems. Thus, fruitful results on this topic have been presented in previous works. In this letter, we analyzed the limitations / deficiency of the original SLNR scheme in multi-layer MIMO systems, and proposed an enhanced SLNR scheme, termed layer SLNR. In the layer SLNR scheme, both the MIMO receiver structure and the inter-layer interference (from the same user and from different users) are formulated in designing the precoder. Therefore, the signal power and interference power are accurately modeled and better balanced. The effectiveness and superiority of the layer SLNR scheme are validated through simulations.